\documentclass[copyright,creativecommons]{eptcs}
\providecommand{\event}{}
\providecommand{\volume}{??}
\providecommand{\anno}{20??}
\providecommand{\firstpage}{1}
\usepackage{breakurl}

\usepackage{amsthm}
\usepackage{amsmath}
\usepackage{amssymb}
\usepackage{mathrsfs}

\title{Characterizing Polynomial Time Computability of Rational and Real Functions}
\author{Walid Gomaa
\institute{INRIA Nancy Grand-Est Research Centre, France}
\institute{Faculty of Engineering, Alexandria University, Alexandria, Egypt}
\email{walid.gomaa@loria.fr}}

\date{}

\def\authorrunning{Walid Gomaa}
\def\titlerunning{Characterizing Polynomial Time Computability of Rational and Real Functions}

\theoremstyle{definition}

\newtheorem{dfn}{Definition}
\newtheorem{lem}[dfn]{Lemma}
\newtheorem{exm}[dfn]{Example}
\newtheorem{thm}[dfn]{Theorem}
\newtheorem{cor}[dfn]{Corollary}

\newtheorem{rem}[dfn]{Remark}
\newtheorem{nota}[dfn]{Notation}

\newcommand{\naturals}{\mathbb{N}}
\newcommand{\reals}{\mathbb{R}}

\newcommand{\dyadics}{\mathbb{D}}

\begin{document}

\maketitle

\begin{abstract}
Recursive analysis was introduced by A. Turing [1936], A. Grzegorczyk [1955],
and D. Lacombe [1955]. It is based on a discrete mechanical framework that can be used to model computation over the real
numbers. In this context the computational complexity of real functions defined over compact
domains has been extensively studied. However, much less have been done for other kinds of real functions. This article is divided into two main parts. The first part investigates polynomial time computability of rational functions and the role of continuity in such computation. On the one hand this is interesting for its own sake. On the other hand it provides insights into polynomial time computability of real functions for the latter, in the sense of recursive analysis, is modeled as approximations of rational computations. The main conclusion of this part is that continuity does not play any role in the efficiency of computing rational functions. The second part defines polynomial time computability of arbitrary real functions, characterizes it, and compares it with the corresponding notion over rational functions. Assuming continuity, the main conclusion is that there is a conceptual difference between polynomial time computation over the rationals and the reals manifested by the fact that there are polynomial time computable rational functions whose extensions to the reals are not polynomial time computable and vice versa.
\end{abstract}

\section{Introduction}

Recursive analysis is an approach for investigating computation over the real numbers; it provides
a theoretical framework for numerical algorithms. The field was introduced by A. Turing \cite{TComputable}, A. Grzegorczyk \cite{GComputable},
and D. Lacombe \cite{LExtension}. The computation model is based on the mechanistic view of
classical computability theory. Hence, unlike other models of real computation
such other C. Moore's recursive class \cite{MRecursion}, recursive analysis takes more direct and realistic approach by appealing
to the notion of \emph{Turing machine} and hence physical realizability. A
conventional machine equipped with function oracles, called \emph{Type II Turing machine}, is adopted.

In this article we assume the binary alphabet $\{0,1\}$. Hence, in such a context finite strings are interpreted as those rationals with finite binary representation. These are called the \emph{dyadic rationals}. We denote them $\dyadics$, and will be assumed throughout the rest of this article as representations of real numbers as indicated by the following definition.

\begin{dfn}(Cauchy sequence representation of real numbers)
Assume $x\in\reals$. Then $x$ can be represented by a Cauchy function
$\varphi_x\colon\naturals\rightarrow\dyadics$ that converges at a binary rate:
\begin{equation}
\forall n\in\naturals\colon|x-\varphi_x(n)|\le2^{-n}
\end{equation}
Given $x\in\reals$, let $CF_x$ denote the class of Cauchy functions that represent $x$.
\end{dfn}

The computational complexity of real functions defined over compact domains has been extensively studied (see for example, \cite{KComplexity}).
However, much less have been done regarding functions over non-compact domains (see foundational work \cite{GComputable,KEgyzzeru,LExtension}, work investigating the
elementary functions and the Grzegorczyk hierarchy \cite{CComputational,CMCAnalog,CMCAnalog_2}, connection with the $GPAC$ model \cite{CMCIteration}, characterization by function algebras \cite{BHRecursive}).
The main crux of the current work is to investigate polynomial time computability of rational and real functions defined over domains with components of the form  $[a,\infty)$.

The article is divided into two main parts. The first part investigates polynomial time computability of rational functions and the role of continuity in such computation. On the one hand this is interesting for its own sake. On the other hand it provides insights into polynomial time computability of real functions for the latter, in the sense of recursive analysis, is modeled as approximations of rational computations. The main conclusion of this part is that continuity does not play any role in the efficiency of computing rational functions. There are polynomial time computable rational functions that are discontinuous. Even we can find efficiently computable functions that are ill-behaved from the smoothness perspective, that is, they have arbitrarily large moduli.

The second part defines polynomial time computability of arbitrary real functions, characterizes it for the particular case where the functions are defined
over domains with components of the form $[a,\infty)$, and finally  compares this notion of computability with the corresponding one over rational functions. Assuming continuity, the main conclusion of this part is that there is a major conceptual difference between polynomial time computation over the rationals and over the reals. This is manifested by the fact that there are continuous polynomial time computable rational functions whose extensions to the reals are not polynomial time computable and conversely, there are rational-preserving polynomial time computable real functions whose restriction to the rationals are not polynomial time computable.

The article is organized as follows. Section 1 is an introduction.
Section 2 gives the basic definitions and notions of polynomial time
computation over $\dyadics$ along with some results about the features of rational computation. Section 3 gives the basic definitions and notions of polynomial time computation over $\reals$
along with some results about the features of real computation and its relationship to rational computation. Section 4 outlines future research directions to be pursued.

\section{Polynomial Time Computability over the Dyadic Numbers}

Let $\Sigma=\{0,1,-,.\}$ and $\Gamma=\{00,01,10,11\}$.
Define a function $\tau\colon\Sigma\to\Gamma$ as follows: $\tau(0)=00,\tau(1)
=11,\tau(-)=01,\tau(.)=10$. Assume that $\mathbb{D}$ is the set of dyadic numbers represented
in lowest forms with the alphabet $\Sigma$.
Define a function $\tau^*\colon\mathbb{D}\to\Gamma^*$
as follows: $\tau^*(a_0,\ldots,a_n)=\tau(a_0)\cdots\tau(a_n)$. For any $d\in\dyadics$ let $len(d)$
denote the length of the binary string $\tau^*(d)$.

\begin{dfn}[$PTime$ (Polynomial time) complexity over $\mathbb{D}$]
Assume a function $f\colon\mathbb{D}\to\mathbb{D}$.
We say that $f$ is \emph{polynomial time computable} if there exists a Turing machine $M$
such that for every $d\in\mathbb{D}$ the following holds
\begin{equation}
M(\tau^*(d))=\tau^*(f(d))
\end{equation}
and the computation time of $M$ is bounded by $p(|\tau^*(d)|)$ for some polynomial function $p$.
\end{dfn}

Let $\mathcal{P}_\mathbb{D}$ denote the class of $PTime$ computable dyadic functions.
For simplicity in the following discussion we focus for the most part on unary functions.

For ease of readability we will use the notations for intervals
to indicate just the dyadics in these intervals. In the following discussion we will need to distinguish
between two subclasses of $\mathcal{P}_\mathbb{D}$. The first class is $\mathcal{P}_\mathbb{D}[\mathscr{B}]=\{f\colon [a,\infty)
\to\mathbb{D}|f\in\mathcal{P}_\mathbb{D},a\in\mathbb{D}\}$. The second class is
$\mathcal{P}_\mathbb{D}[\mathscr{U}]=\{f\colon (a,\infty)
\to\mathbb{D}|f\in\mathcal{P}_\mathbb{D},a\in\mathbb{D}\}\cup\{f\colon \mathcal{D}
\to\mathbb{D}| \mathcal{D}=(a,b)\;or\;\mathcal{D}=(a,b]\;or\;\mathcal{D}=[a,b),f\in\mathcal{P}_\mathbb{D};a,b\in\mathbb{D}\}$.
By removing the restriction $f\in\mathcal{P}_\mathbb{D}$, let $\mathbb{D}[\mathscr{B}]$ and $\mathbb{D}[\mathscr{U}]$
denote the resulting classes.

\begin{dfn}[Continuous dyadic functions]~
\label{dfn:continuous_dyadic_functions}
\begin{enumerate}
\item{Assume a function $f\in\mathcal{P}_\mathbb{D}[\mathscr{B}]$ with domain $[a,\infty)$. We say that $f$ is \emph{continuous}
if $f$ has a \emph{modulus of continuity}, that is if there exists a function $m\colon\mathbb{N}^2\to\mathbb{N}$
such that for every $k,n\in\mathbb{N}$ and for every $x,y\in[a,a+2^k]$ the following
holds:
\begin{equation}
if\; |x-y|\le 2^{-m(k,n)}, then\; |f(x)-f(y)|\le 2^{-n}
\end{equation}}
\item{Assume a function $f\in\mathcal{P}_\mathbb{D}[\mathscr{U}]$ with domain $(a,b)$ (other cases can be similarly
handled). We say that $f$ is \emph{continuous}
if $f$ has a \emph{modulus of continuity}, that is if there exists a function $m\colon\mathbb{N}^2\to\mathbb{N}$
such that for every $k,n\in\mathbb{N}$ and for every $x,y\in[a+2^{-k},b-2^{-k}]$ the following
holds:
\begin{equation}
if\; |x-y|\le 2^{-m(k,n)}, then\; |f(x)-f(y)|\le 2^{-n}
\end{equation}}
\end{enumerate}
\end{dfn}

In the following we will refer to $k$ as the \emph{extension argument} and $n$ as the \emph{precision argument}.

\begin{rem}
\begin{enumerate}
\item{It is clear from Definition \ref{dfn:continuous_dyadic_functions} that the open domain of a function
is divided into a sequence of compact subintervals, each of which corresponds to a fixed $k$.
In the following discussion we will typically need to fix such a compact subinterval, hence to ease the notation
we use $m_k(n)=m(k,n)$ to indicate the modulus of continuity over the fixed subinterval.}
\item{It is also clear that any continuous dyadic function
is bounded over any compact subinterval.}
\item{As opposed to the case of $\mathbb{R}$-computation, computability
over $\mathbb{D}$ does not imply continuity of the underlying function.}
\end{enumerate}
\end{rem}

Let $\mathcal{P}_\mathbb{D}[cnt]$ denote the continuous subclass of
$\mathcal{P}_\mathbb{D}$, similarly for $\mathcal{P}_\mathbb{D}[\mathscr{B},cnt]$ and
$\mathcal{P}_\mathbb{D}[\mathscr{U},cnt]$. Let $\mathcal{P}_\mathbb{N}$ denote the
class of $PTime$ computable $\mathbb{N}$-functions. Assume a unary function $f\in\mathcal{P}_\mathbb{N}$.
Let $\tilde{f}
\in\mathcal{P}_\mathbb{D}[\mathscr{B}]$ with domain $[0,\infty)$ be an extension of $f$ defined as follows:
\begin{equation}
\tilde{f}(x)=f(\lfloor x\rfloor)
\end{equation}
Let $\tilde{\mathcal{P}}=\{\tilde{f}\colon f\in\mathcal{P}_\mathbb{N}\}$ and let $\dyadics_0^+$ denote the
set of nonnegative dyadics.

\begin{lem}
\label{lem:PTime_dyadic_functions_are_not_bounded_by_PTime_natural_functions}
There exists $f\in\mathcal{P}_\mathbb{D}[\mathscr{B},cnt]$ whose computation time is not bounded by any function
in $\tilde{\mathcal{P}}$.
In other words for every $\tilde{f}\in\tilde{\mathcal{P}}$ the following
holds for infinitely many $d\in dom(f)$: $len(f(d)) > \tilde{f}(d)$.
\end{lem}
\begin{proof}
We will construct a function $f$ such that through an interval $[r,r+1]$, for $r\in\naturals$, the
function grows piecewise linear with predetermined breakpoints chosen such that the length
of $f$ grows polynomially in terms of the length of the dyadic input, however, it grows exponentially
in terms of the length of the integer part of the input. For $a\in\mathbb{N}$, let $\epsilon_a$ denote the fraction $0.(01)^a$.
Let $d_0=r$, and let $k=\min\{i\in\mathbb{N}\colon r+1\in[0,2^i]\}$.
For every $j\in\{1,\ldots,r\}$ let $d_j=d_0+\epsilon_j$,
$\delta_j=\epsilon_j-\epsilon_{j-1}=2^{-2j}$. The $d_j$'s are breakpoints through which the function
increases piecewise linearly. Let $e_0=d_r+\epsilon_r$ and for every $j\in\{1,\ldots,r\}$
let $e_j=e_0-\epsilon_j$. The $e_j$'s are breakpoints through which the function decreases piecewise linearly, these
are needed to maintain continuity.
The formal definition of $f\colon\mathbb{D}_0^+\to\mathbb{D}$ (over every interval $[r,r+1]$) is as follows:
\begin{equation}
f(d)=
\begin{cases}
0 & d\in\mathbb{N}\\
0 & e_0\le d\le r+1\\
2^j & d=d_j,j\in\{1,\ldots,r\}\\
2^j & d=e_j,j\in\{1,\ldots,r\}\\
\frac{\delta f(d_{j+1})+(\delta_{j+1}-\delta)f(d_j)}{\delta_{j+1}} & d_j<d<d_{j+1},\delta=d-d_j\\
\frac{\delta f(e_{j+1})+(\delta_{j+1}-\delta)f(e_j)}{\delta_{j+1}} & e_{j+1}<d<e_j,\delta=e_j-d
\end{cases}
\end{equation}
Note that $len(f(d_j))$ is linear in $len(d_j)$, similarly for $len(f(e_j))$,
hence $f$ is $PTime$ computable.
Note that $f(d_r)=2^r=\Omega(2^{2^{len(r)}})$, hence $f$ is not $PTime$
computable with respect to $\naturals$-points and therefore its computation is not bounded by any function in $\tilde{\mathcal{P}}$.
By its definition $f$ is continuous.
In the following we will find a modulus function for $f$.
Define a function $m\colon\mathbb{N}^2\to\mathbb{N}$
by $(k,n)\mapsto 3\cdot2^k+n$. Let $\ell_j$
denote the interval $[d_{j-1},d_j]$ for $j\in\{1,\ldots,r\}$.
Assume
$x,y\in[r,d_r]$ such that $|x-y|\le 2^{-m(k,n)}$ (other cases are either
trivial or can be handled symmetrically).
Note that $f$ is monotonically increasing piecewise linear over $[r,d_r]$ and
the slope of the line in interval $\ell_j$, where $j>1$, can be computed as follows:
\begin{align*}
f'_j&=\frac{f(d_j)-f(d_{j-1})}{\delta_{j}}\\
& = \frac{2^j-2^{j-1}}{2^{-2j}}\\
& = 2^{3j-1}
\end{align*}
\noindent\underline{case 1:} $x,y\in\ell_j$: Note that $\delta_j=2^{-2j}\ge\delta_r=2^{-2r}$.
We have
\begin{align*}
|f(y)-f(x)|&= |(y-x)\frac{f(y)-f(x)}{y-x}|\\
&=|y-x|f'_j\\
&\le |y-x|f'_r\\
&\le 2^{-(3\cdot2^k+n)}f'_r\\
&\le 2^{-(3\cdot2^k+n)}2^{3r-1}\\
&\le2^{-(n+1)}
\end{align*}
\noindent\underline{case 2:} $x\in\ell_j$ and $y\in\ell_{j+1}$ where $1<j\le r-1$:
\begin{align*}
|f(x)-f(y)|&\le|f(x)-f(d_j)|+|f(d_j)-f(y)|\\
&\le 2^{-(n+1)}+2^{-(n+1)},\qquad \text{(from case 1)}\\
&= 2^{-n}
\end{align*}
The total length of the two smallest intervals $\ell_r,\ell_{r-1}$ is $\delta_r+\delta_{r-1}=2^{-2r}+2^{-2(r-1)}
> 2^{-2(r-1)}\ge 2^{-(3r+n)}\ge |x-y|$.
Hence, it can not happen that $x\in\ell_i$
and $y\in\ell_j$ with $|j-i|>1$. Therefore, $m$ is a modulus function for $f$.
\end{proof}

\begin{lem}
\label{lem:PTime_dyadic_functions_may_have_modulus_not_poly_in_k}
There exists $f\in\mathcal{P}_\mathbb{D}[\mathscr{B},cnt]$
that does not have a polynomial modulus with respect to the extension argument $k$.
\end{lem}
\begin{proof}
Investigating the proof of Lemma \ref{lem:PTime_dyadic_functions_are_not_bounded_by_PTime_natural_functions},
it can be easily seen that the function $f$ defined in that proof satisfies the conclusion of this lemma.
\end{proof}

\begin{rem}
By looking again into the proof of Lemma \ref{lem:PTime_dyadic_functions_are_not_bounded_by_PTime_natural_functions},
we observe that the apex of the graph of $f$ can be taken as arbitrarily high as we want by letting $j$ runs from $1$ to
$\alpha(r)$ for any monotonically increasing function $\alpha$. This indicates that there is no upper bound
on the moduli of continuity of the functions in $\mathcal{P}_\mathbb{D}[\mathscr{B},cnt]$.
\end{rem}

\begin{lem}
\label{lem:PTime_dyadic_functions_may_have_modulus_not_poly_in_n}
There exists a function $f\in\mathcal{P}_\mathbb{D}[\mathscr{B},cnt]$ that
does not have a polynomial modulus with respect to the precision argument $n$.
\end{lem}
\begin{proof}
Define a function $\alpha\colon\mathbb{N}\to\mathbb{N}$ by
\begin{equation}
\alpha(i)=
\begin{cases}
0 & i=0\;or\;i=1\\
\max\{j\le i\colon \lfloor\log_2{\log_2{j}}\rfloor=\log_2{\log_2{j}}\} & ow
\end{cases}
\end{equation}
For every $i\in\mathbb{N}$ let $d_i=1-2^{-i}$. Define a function $f\colon\mathbb{D}_0^+\to\mathbb{D}$ as follows:
\begin{equation}
f(d)=
\begin{cases}
1 & d_0\le d\le d_1\\
\frac{1}{\log_2{\alpha(i)}} & d=d_i,i>1\\
2^{(i+1)}\left(\delta f(d_{i+1})+(2^{-(i+1)}-\delta)f(d_i)\right) & d_i<d<d_{i+1}, \delta=d-d_i\\
0 & d\ge 1
\end{cases}
\end{equation}

Then $f$ is a piecewise linear decreasing function over the interval $[0,1]$.
It decreases very slowly (may even remain constant
over long subintervals), however, it eventually reaches $0$ at $d=1$, thus it is continuous.
Note that for every $i$, $len(d_i)=i$ and by definition $len(f(d_i))=O(\log(i))$. In addition
$f(d_i)$ is efficiently computable. Hence, $f$ is $PTime$ computable.
Finally, we need to show that $f$ does not have a polynomial modulus
with respect to the precision parameter.
Let $\ell_i$ denote the subinterval $[d_{i-1},d_i]$. Note that there are infinitely many $\ell_i$'s
over which $f$ is strictly decreasing.
The goal is to compute the slope of the function over such subintervals.
Assume such an interval $\ell_i=[d_{i-1},d_i]$,
then it must be the case that $i=2^{2^j}$ for some $j\in\mathbb{N}$.
\begin{align*}
|\ell_i|&=d_i-d_{i-1}\\
&=1-2^{-i}-1+2^{-(i-1)}\\
&=2^{-i}=2^{-2^{2^j}}
\end{align*}
On the other hand
\begin{align*}
f(d_{i-1})-f(d_i)&=2^{-(j-1)}-2^{-j}\\
&=2^{-j}
\end{align*}
Hence, the slope of the line over $\ell_i$ is
\begin{align*}
|f'_i|&=\frac{2^{-j}}{2^{-2^{2^j}}}\\
&=2^{2^{2^j}-j}
\end{align*}
which can not be captured by any polynomial function.
\end{proof}

Combining the proofs of Lemma \ref{lem:PTime_dyadic_functions_may_have_modulus_not_poly_in_k}
and Lemma \ref{lem:PTime_dyadic_functions_may_have_modulus_not_poly_in_n} we have the following.

\begin{thm}
\label{thm:dyadic_cnt_function_with_nonpoly_modulus_wrt_k_n}
There exists a function $f\in\mathcal{P}_\mathbb{D}[\mathscr{B},cnt]$ that
does not have a polynomial modulus with respect to both the extension parameter $k$ and
the precision parameter $n$ (that is if one variable is considered constant
the function would not be polynomial in the other).
\end{thm}

\section{Polynomial Time Computability over the Real Numbers}

Since continuity is a necessary condition for computation over $\mathbb{R}$,
it will not be mentioned explicitly.
So we use the notation $\mathcal{P}_\mathbb{R}$
to denote continuous $PTime$ computable $\mathbb{R}$-functions.
Again we divide into two subcases:
$\mathcal{P}_\mathbb{R}[\mathscr{B}]$ and $\mathcal{P}_\mathbb{R}[\mathscr{U}]$.
In this section we exclusively handle the case $\mathcal{P}_\mathbb{R}[\mathscr{B}]$. In the case
of real functions over compact domains there is only one parameter that controls the
computational complexity, namely \emph{the precision} (the level of approximation
required for the output). Given a positive integer $n$ as an
input to the machine computing such a function, then roughly speaking $n$ is the required
length of the output. Over a \emph{compact domain} there are predetermined lower and upper bounds on the
function value, hence the length of the integer part can be considered constant and
therefore does not play any role in the complexity. For a detailed discussion about functions
over compact domains see \cite{KComplexity}. Moving to functions in
$\mathcal{P}_\mathbb{R}[\mathscr{B}]$ the domain and generally the range become unbounded hence an additional parameter,
namely the length of the integer part, must now be accounted for in the computational complexity
of a function.
Unlike the precision parameter which is an explicit input the extension parameter is extracted by the
machine by asking the oracle that represents the actual real number input.

Given a Cauchy sequence $\varphi$ let $M^{^{\varphi}}$ denote a Turing machine
$M$ that has access to oracle $\varphi$.

\begin{dfn}($PTime$ complexity over $\mathbb{R}[\mathscr{B}]$)
\label{dfn:Polynomial time computable real functions case_1}
Assume a function $f\colon[a,\infty)\rightarrow\mathbb{R}$.
We say that $f$
is \emph{polynomial time computable} if the following conditions hold:
\begin{enumerate}
\item{There exists a two-function oracle Turing machine such that for every $\varphi_a\in CF_a$, $x\in[a,\infty)$,
$\varphi_x\in CF_x$, and for every $n\in\mathbb{N}$ the following holds:
\begin{equation}
|M^{^{\varphi_x,\varphi_a}}(n) - f(x)|\le 2^{-n}
\end{equation}}
\item{The \emph{computation time} of $M^{^{()}}(n)$ is bounded by $p(k,n)$ for some
polynomial $p$, where $k=\min\{j\colon x\in[a,a+2^j]\}$.}
\end{enumerate}
\end{dfn}

\begin{rem}
Note that in the previous definition there was not any computability restrictions
over the constant $a$. Furthermore, we chose to universally quantify over all
possible Cauchy sequence representations of $a$. This avoids the risk of apriori fixing
a particular Cauchy sequence which might contain hyper computational information (such as the encoding of
the halting set).
\end{rem}

\begin{exm}
Consider the function $f\colon[0,\infty)\rightarrow\mathbb{R}$, defined by $f(x)=e^x$.
Note that $f(x)=\Omega(2^x)$. Since $2^x\upharpoonright\mathbb{N}$ is not polynomial
time computable as an $\mathbb{N}$-function, it is not either polynomial time computable as an $\mathbb{R}$-function.
 Thus, $f$ is
not polynomial time computable.
\end{exm}

\begin{nota}
For any $x\in\mathbb{R}$, let $\varphi_x^*\in CF_x$ denote the particular Cauchy
function
\begin{equation}
\varphi_x^*(n)=\frac{\lfloor 2^n\cdot x \rfloor}{2^n}
\end{equation}
\end{nota}

\begin{thm}(Characterizing $\mathcal{P}_\mathbb{R}[\mathscr{B}]$)
\label{thm:characterizing_ptime_over_reals_case_1}
Assume a function $f\colon[a,\infty)\rightarrow\mathbb{R}$.
Then $f$ is \emph{polynomial time computable} iff there exist two functions
$m:\mathbb{N}^2\rightarrow
\mathbb{N}$ and $\psi:\mathbb{D}\cap[a,\infty)\times\mathbb{N}\rightarrow\mathbb{D}$
such that
\begin{enumerate}
\item{$m$ is a modulus function for $f$ and it is polynomial with respect to both the extension parameter $k$ and
the precision parameter $n$,
that is, $m(k,n)=(k+n)^b$ for some $b\in\mathbb{N}$.}
\item{$\psi$ is an \emph{approximation function} for $f$, that is,
for
every $d\in\mathbb{D}\cap[a,\infty)$ and every $n\in\mathbb{N}$ the following holds:
\begin{equation}
|\psi(d,n)-f(d)|\le 2^{-n}
\end{equation}}
\item{$\psi(d,n)$ is computable in time $p(|d|+n)$ for some polynomial $p$.}
\end{enumerate}
\end{thm}
\begin{proof}
Fix some $\varphi_a\in CF_a$. The proof is an extension of the proof of Corollary 2.21 in \cite{KComplexity}.
Assume the existence of $m$ and $\psi$ that satisfy the given conditions.
Assume an $f$-input $x\in[a,\infty)$ and let $\varphi_x\in CF_x$. Assume $n\in\naturals$.
Let $M^{^{\varphi_x,\varphi_a}}(n)$ be an oracle Turing machine that does the following:

\begin{enumerate}
\item{let $d_1=\varphi_a(2)$ and $d_2=\varphi_x(2)$,}
\item{from $d_1$ and $d_2$ determine the least $k$ such that $x\in[a,a+2^k]$,}
\item{let $\alpha=m(k,n+1)$,}
\item{let $d=\varphi_x(\alpha)$,}
\item{let $e=\psi(d,n+1)$ and output $e$.}
\end{enumerate}


Note that every step of the above procedure can be performed in polynomial time
with respect to both $k$ and $n$.
Now
verifying the correctness of $M^{^{()}}(n)$:
\begin{align*}
|e-f(x)|&\le|e-f(d)|+|f(d)-f(x)|\\
&\le 2^{-(n+1)}+|f(d)-f(x)|,\qquad \mbox{by definition of $\psi$}\\
& \le 2^{-(n+1)}+2^{-(n+1)},\qquad \mbox{$|d-x|\le 2^{-m_k(n+1)}$ and definition of $m$}\\
& =2^{-n}
\end{align*}
This completes the first part of the proof.  Now assume $f$ is polynomial time computable. We adopt the following notation: for every $x\in\reals$ let $\varphi_x^*(n)=\frac{\lfloor 2^nx\rfloor}{2^n}$.
Fix some large enough $k$ and consider any $x\in[a,\infty)$ such that $len(\lfloor x\rfloor)=k+len(\lfloor a\rfloor)$, hence
$x\in[a,a+2^k]$. For simplicity in the following discussion we will ignore the constant $len(\lfloor a\rfloor)$. Since
$f$ is polytime computable, there exists an oracle Turing machine $M^{^{()}}$ such that the computation time of
$M^{^{\varphi_x^*,\varphi_a}}(n)$ is bounded by $q(k,n)$ for some polynomial $q$.
Fix some large enough $n\in\mathbb{N}$.

Let
\begin{equation}
n_x=\max\{j\colon\varphi_x^*(j)\textit{ is queried during the computation of }
M^{^{\varphi_x^*,\varphi_a}}(n+3)\}
\end{equation}

Let $d_x=\varphi_x^*(n_x)$. By the particular choice of Cauchy sequences we have $\varphi_{d_x}^*(j)=\varphi_x^*(j)$ for every $j\le n_x$.
Let $\ell_x=d_x-2^{-n_x}$ and $r_x=d_x+2^{-n_x}$. Then $\{(\ell_x,r_x)\colon x\in[a,a+2^k]\}$ is an \emph{open covering} of
the compact interval $[a,a+2^k]$. By the \emph{Heine-Borel Theorem}, $[a,a+2^k]$ has a finite covering
$\mathcal{C}=\{(\ell_{x_i},r_{x_i})\colon i=1,\ldots,w\}$. Define
$m'\colon\mathbb{N}^2\to\mathbb{N}$ by

\begin{equation}
m'(k,n)=\max\{n_{x_i}\colon i=1,\ldots,w\}
\end{equation}

First We need to show that $m'$ is a modulus for $f$. Assume some $x,y\in[a,a+2^k]$ such that $x<y$ and $|x-y|\le 2^{-m'_k(n)}$.\\
\noindent\underline{case 1:} $x,y\in(\ell_{x_i},r_{x_i})$ for some $i\in\{1,\ldots,w\}$.
Then $|x-d_{x_i}|< 2^{-n_{x_i}}$ which implies that $\varphi_x^*(j)=\varphi_{x_i}^*(j)=\varphi_{d_{x_i}}^*(j)$ for every $j\le n_{x_i}$,
hence $M^{^{\varphi_x^*}}(n+3)=M^{^{\varphi_{x_i}^*}}(n+3)=M^{^{\varphi_{d_{x_i}}^*}}(n+3)$.
Now
\begin{align*}
|f(x)-f(d_{x_i})|&\le|f(x)-M^{^{\varphi_x^*}}(n+3)|+|M^{^{\varphi_x^*}}(n+3)-f(d_{x_i})|\\
&=|f(x)-M^{^{\varphi_x^*}}(n+3)|+|M^{^{\varphi_{d_{x_i}}^*}}(n+3)-f(d_{x_i})|\\
&\le 2^{-(n+3)}+2^{-(n+3)}\\
&=2^{-(n+2)}
\end{align*}
Similarly, we can deduce that $|f(y)-f(d_{x_i})|\le 2^{-(n+2)}$. Hence, $|f(x)-f(y)|\le|f(x)-f(d_{x_i})|+|f(d_{x_i})-f(y)|
\le 2^{-(n+2)}+2^{-(n+2)}=2^{-(n+1)}$.\\
\noindent\underline{case 2:} There is no $i$ such that $x,y\in(\ell_{x_i},r_{x_i})$. Notice that $\mathcal{C}$
is a covering and by assumption $|x-y|\le\min\{\frac{1}{2}(r_{x_i}-\ell_{x_i})\colon i=1,\ldots,w\}$.
Hence there must exist $i,j$ such that
$x\in(\ell_{x_i},r_{x_i})$, $y\in(\ell_{x_j},r_{x_j})$, and $\ell_{x_j}<r_{x_i}$.
Choose an
arbitrary $z\in(\ell_{x_j},r_{x_i})$. Then
\begin{align*}
|f(x)-f(y)|&\le |f(x)-f(z)|+|f(z)-f(y)|\\
& \le 2^{-(n+1)}+|f(z)-f(y)|,\quad\quad\textit{applying case 1 to $x,z\in(\ell_{x_i},r_{x_i})$}\\
&\le 2^{-(n+1)}+2^{-(n+1)},\quad\quad\textit{applying case 1 to $y,z\in(\ell_{x_j},r_{x_j})$}\\
&=2^{-n}
\end{align*}
Hence, $m'$ is a modulus function for $f$.
From the assumption on the time complexity of $f$ we have
$n_x\le q(k,n+3)$. Hence, the function $m(k,n)=q(k,n+3)$ is a modulus function for $f$.
The approximation function can be defined as follows:
for $d\in\dyadics$ and $n\in\naturals$, let $\psi(d,n)=M^{^{\varphi_d^*,\varphi_a}}(n)$.
This completes the proof of the theorem.
\end{proof}

The following result is a consequence of Theorem \ref{thm:dyadic_cnt_function_with_nonpoly_modulus_wrt_k_n}
and Theorem \ref{thm:characterizing_ptime_over_reals_case_1}.

\begin{thm}
\label{thm:dyadic_fun_ptime_its_real_extension_is_not}
There exists a dyadic
continuous function that is $PTime$ computable, however, its extension to
$\mathbb{R}$ is not $PTime$ computable.
\end{thm}

Let $\mathcal{P}_\mathbb{R}[\mathscr{B}]=\{f\in\mathbb{R}[\mathscr{B}]\colon f\;is\;PTime\;computable\}$
and let $\mathcal{P}_\mathbb{D}[\mathscr{B},cnt,poly]=\{f\in\mathcal{P}_\mathbb{D}[\mathscr{B},cnt]\colon\\ f
\;has\;a\;polynomial\;modulus\;with\;respect\;to\;both\;arguments\}$.
Let $\mathcal{P}_\mathbb{R}[\mathscr{B}]
\upharpoonright\mathbb{D}=\{f\in\mathbb{D}[\mathscr{B}]\colon\exists\tilde{f}\in\mathcal{P}_\mathbb{R}[\mathscr{B}]
\;such\;that\;f=\tilde{f}\upharpoonright\mathbb{D}\}$.
The following result shows the converse of Theorem
\ref{thm:dyadic_fun_ptime_its_real_extension_is_not},
that is there exists a dyadic-preserving real function that
is $PTime$ computable, however, its restriction to $\mathbb{D}$ is not $PTime$
computable.

\begin{thm}
\label{thm:ptime_as_real_but_not_as_dyadic}
There exists a function $f\in\mathcal{P}_\mathbb{R}[\mathscr{B}]
\upharpoonright\mathbb{D}$ that is not $PTime$ computable.
\end{thm}
\begin{proof}
Define a function $f\colon[0,\infty)\to\mathbb{R}$ as follows:
\begin{equation}
f(x)=
\begin{cases}
0 & x\in\mathbb{N}\\
\frac{1}{2}+2^{-2^k}& x=j+\frac{1}{2}\;for j\in\mathbb{N}\;and\;k=\min\{i\in\mathbb{N}\colon x<2^i\}\\
2(x-j) f(j+\frac{1}{2}) & j<x<j+\frac{1}{2}, j\in\mathbb{N}\\
2(j+1-x)f(j+\frac{1}{2}) & j+\frac{1}{2}<x<j+1,j\in\mathbb{N}
\end{cases}
\end{equation}

$f$ is piecewise linear with breakpoints at $j$'s and $(j+\frac{1}{2})$'s for $j\in\naturals$. It is
$zero$ at the integer points and $\frac{1}{2}+\epsilon_j$ at the midpoints where $\epsilon_j$ is a very small
value that depends on the binary length of $j$. The idea is that real computation is inherently
approximate hence to get the exact correct value at $j+\frac{1}{2}$ the precision input $n$ has to be large
enough (much larger than the extension parameter)
making the complexity polynomial in terms of $n$ although it is exponential in terms of the extension parameter.
And therefore the overall complexity is polynomial.
On the other hand dyadic computation does not involve this precision parameter leaving the overall
computation exponential in terms of the only remaining extension parameter. Now we give the technical details.
It is clear that $f$ preserves $\mathbb{D}$. Let $g=f\upharpoonright\mathbb{D}$.
Assume some
$x\in dom(g)$ such that $x=j+\frac{1}{2}$ for some $j\in\mathbb{N}$. Let $k=len(j)$.
From the definition of $f$, $len(g(x))=\Omega(2^k)$. Hence $g$ is not $PTime$ computable as a dyadic
function. Now remains to show $f$ is $PTime$ computable as a real function. Assume some $x\in dom(f)$
and assume some $\varphi\in CF_x$. Let $M^{^{()}}$ be an oracle Turing machine such that
$M^{^\varphi}(n)$ does the following:
\begin{enumerate}
\item{Let $e=\varphi(2)$,}
\item{Determine the least $k$ such that $e+1<2^k$,}
\item{Let $d=\varphi(n+3)$,}
\item{If $j\le d\le j+\frac{1}{2}$ for some $j\in\mathbb{N}$, then
\begin{enumerate}
\item{If $n\ge 2^k-10$, then output $2(d-j)(\frac{1}{2}+2^{-2^k})$,}
\item{else output $(d-j)$,}
\end{enumerate}}
\item{If $j+\frac{1}{2}\le d\le j+1$ for some $j\in\mathbb{N}$, then
\begin{enumerate}
\item{If $n\ge 2^k-10$, then output $2(j+1-d)(\frac{1}{2}+2^{-2^k})$,}
\item{else output $(j+1-d)$,}
\end{enumerate}}
\item{End.}
\end{enumerate}
Clearly $M^{^\varphi}(n)$ runs in polynomial time with respect to $n$ and $k$.
We need to
show its correctness. Assume $\epsilon=|x-d|\le2^{-(n+3)}$. We have the following cases.\\
\noindent\underline{case 1:} $x,d\in[j,j+\frac{1}{2}]$. If $n\ge 2^k-10$, then
\begin{align*}
|M^{^\varphi}(n)-f(x)|&=|2(d-j)f(j+\frac{1}{2})-2(x-j)f(j+\frac{1}{2})|\\
&=2f(j+\frac{1}{2})|x-d|\\
&\le 2(\frac{1}{2}+2^{-2^k})2^{-(n+3)}\\
&=2^{-(n+3)}+2^{-(2^k+n+2)}\\
&\le 2^{-(n+2)}
\end{align*}

If $n<2^k-10$, then
\begin{align*}
|M^{^\varphi}(n)-f(x)|&=|(d-j)-2(x-j)(\frac{1}{2}+2^{-2^k})|\\
&=2|\frac{1}{2}(d-j)-(x-j)(\frac{1}{2}+2^{-2^k})|\\
&=2|\frac{1}{2}d-\frac{1}{2}j-\frac{1}{2}x+\frac{1}{2}j-2^{-2^k}(x-j)|\\
&=2|\frac{1}{2}(d-x)-2^{-2^k}(x-j)|\\
&\le2(|\frac{1}{2}(d-x)|+|2^{-2^k}(x-j)|)\\
&\le 2^{-(n+3)}+2^{-2^k}\\
&\le 2^{-(n+2)}
\end{align*}

\noindent\underline{case 2:} $x,d\in[j+\frac{1}{2},j+1]$. This case is symmetrical with case 1.\\
\noindent\underline{case 3:} One of $x$ or $d$ is in $[j,j+\frac{1}{2}]$ and the other is in
$[j+\frac{1}{2},j+1]$.
Then
\begin{align*}
|M^{^\varphi}(n)-f(x)|&\le|M^{^\varphi}(n)-f(j+\frac{1}{2})|+|f(j+\frac{1}{2})-f(x)|\\
&\le2^{-(n+2)}+2^{-(n+2)},\qquad\textit{from previous cases}\\
&=2^{-(n+1)}
\end{align*}

Similar calculations for the case when either $x$ or $d$ is in $[j+\frac{1}{2},j+1]$ and the other in
$[j+1,j+\frac{3}{2}]$ with $f(j+\frac{1}{2})$ replaced by $f(j+1)$.\\

Hence, $f$ is $PTime$ computable as a real function and this completes the proof of the lemma.
\end{proof}

Theorem \ref{thm:dyadic_fun_ptime_its_real_extension_is_not} and Theorem \ref{thm:ptime_as_real_but_not_as_dyadic}
lead to the following interesting surprising corollary.

\begin{cor}
There exists $f\in\mathcal{P}_\mathbb{D}[\mathscr{B},cnt]$ such that its extension to $\mathbb{R}$
is not polynomial time computable. And there exists a dyadic-preserving real function $g\in\mathcal{P}_{\mathbb{R}}[\mathscr{B}]$
such that $g\upharpoonright\mathbb{D}$ is not polynomial time computable.
\end{cor}

This corollary basically states that polynomial time complexity over the reals is not simply
an extension of the corresponding notion over the dyadics; this is in spite of the fact that real computation
is approximated (in the sense of recursive analysis) by dyadic computation.
This can be justified by the following observations:
(1) the notion of modulus of continuity does not play any role in the computation
of dyadic functions, there even exist efficiently computable dyadic
functions that do not have modulus of continuity, so computation of a dyadic function is not
related, or at most weakly related, to the smoothness of the function, (2) on the contrary
continuity of real functions is a necessary condition for their computability,
(3)
there are two factors controlling the complexity of computing a dyadic function (and finite objects in general):
how hard it is to compute every single bit of the output and the length of the output, and
(4) on the other hand there are three factors
controlling the complexity of computing a real function (in the sense of recursive analysis);
(i) the first, same as in the dyadic case, is how hard it is to compute every single bit of the output,
(ii) the second, partially similar to the dyadic case, is the length of the integer part of the output
(the length of the fractional part is already controlled by the required precision
which is already an input to the machine),
and
(iii) the third factor (and this is the one absent from the dyadic case) is how hard it is to access the input and this
is essentially controlled by the modulus function.

\section{Further Research Directions}

We intend to pursue the following lines of thought:

\begin{enumerate}
\item{Investigate polynomial time computability of rational and real functions that are defined over non-compact domains of the form $\mathscr{U}$.}
\item{Characterize the classes $\mathcal{P}_\dyadics[\mathscr{B},cnt]$ and $\mathcal{P}_\dyadics[\mathscr{B},discnt]$ by function algebras. A candidate function algebra (for the continuous case) would be an
extension to the dyadics of the Bellantoni and Cook class \cite{BCNew}.}
\item{Characterize the classes $\mathcal{P}_\dyadics[\mathscr{U},cnt]$ and $\mathcal{P}_\dyadics[\mathscr{U},discnt]$ by function algebras.}
\item{Characterize the classes $\mathcal{P}_\reals[\mathscr{B}]$, $\mathcal{P}_\reals[\mathscr{U}]$,
and $\mathcal{P}_\reals$ by function algebras.}
\end{enumerate}

\bibliographystyle{eptcs}
\bibliography{refs}

\end{document}